# Fabrication and Low Temperature Thermoelectric Properties of Na$_x$CoO$_2$ ($x$ = 0.68 and 0.75) Epitaxial Films by the Reactive Solid-Phase Epitaxy

W. J. Chang*, C. C. Hsieh, T. Y. Chung, S. Y. Hsu, K. H. Wu, and T. M. Uen

*Department of Electrophysics, National Chiao Tung University, Hsinchu 30010, Taiwan*

J.-Y. Lin and J. J. Lin

*Institute of Physics, National Chiao Tung University, Hsinchu 30010, Taiwan*

C.-H. Hsu

*National Synchrotron Radiation Research Center (NSRRC), Hsinchu 30076, Taiwan*

Y. K. Kuo

*Department of Physics, National Dong Hua University, Hualien 97401, Taiwan*

H. L. Liu, M. H. Hsu, Y. S. Gou, and J. Y. Juang*[†]

*Department of Physics, National Taiwan Normal University, Taipei 10610, Taiwan*

We have fabricated Na$_x$CoO$_2$ thin films via lateral diffusion of sodium into Co$_3$O$_4$ (111) epitaxial films (reactive solid-phase epitaxy: Ref. 4). The environment of thermal diffusion is key to the control of the sodium content in thin films. From the results of x-ray diffraction and in-plane $\rho_{ab}$, the epitaxial growth and the sodium contents of these films were identified. The thermoelectric measurements show a large thermoelectric power similar to that observed in single crystals. The quasiparticle scattering rate is found to approach zero at low temperatures, consistent with the small residual resistivity, indicating high quality of the Na$_x$CoO$_2$ thin films.

* Electronic mail: wei.ep90g@nctu.edu.tw & jyjuang@cc.nctu.edu.tw
[†] Also at the Department of Electrophysics, National Chiao Tung University.

The research on sodium cobaltate $Na_xCoO_2$ (NCO) has led to observations of many unusual physical properties exhibited in related compounds, such as the thermoelectric power,[1] charge ordered insulator at $x = 0.5$,[2] and antiferromagnetic metal at $x = 0.75$.[2] Furthermore, the discovery of superconductivity in $Na_xCoO_2 \cdot yH_2O$ was indeed a surprise.[3] Many theoretical and experimental works have focused on the fascinating and yet puzzling ground states of NCO due to its two-dimensional triangular lattice and the mixed valence character. For the purpose of research and applications, scientists have been trying to fabricate NCO in its thin film form with modulated sodium concentrations, $x$. Owing to the high vapor pressure of Na, however, it is difficult to directly grow high-quality epitaxial NCO thin films by pulsed-laser deposition, not to mention controlling $x$. Very recently, reactive solid-phase growth via lateral diffusion of sodium was reported as a viable method of growing epitaxial NCO thin films.[4] However, precise control of sodium content and related characterizations remained to be resolved. In this letter, we reported a capped-sandwich lateral diffusion scheme capable of fabricating high quality epitaxial NCO films with well-controlled sodium contents.

In order to overcome the high vapor pressure of sodium and its associated detrimental consequences in the firsthand growth of NCO thin films, we divided the procedures of growing epitaxial NCO-(0001) films into two parts. First, a layer of ~

125 nm thick $Co_3O_4$-(111) thin film was deposited on sapphire-(0001) substrates by pulsed-laser deposition (KrF excimer laser, $\lambda$ = 248 nm) with $T_{substrate}$ = 680ºC, $P_{oxygen}$ = 0.2 Torr. Subsequently, thermal diffusion sodium was deployed to the obtained $Co_3O_4$-(111) to intercalate sodium into the $CoO_2$ layers. In our scheme, in order to keep the film surface clean and smooth during the sodium diffusion process, the $Co_3O_4$-(111) film was capped with a sapphire sheet as a sandwich arrangement. This fabrication process has been developed by Ohta et al..[4] And the process is named the reactive solid-phase epitaxy (R-SPE). For more details, see Ref. 5. In order to fabricate $Na_{0.68}CoO_2$ (NCO068) film, we used sodium carbonate powder to muffle the "sandwiched" $Co_3O_4$-(111) film and pressed the whole assembly into pellet, depicted as specimen A in Fig. 1. The lateral diffusion process was carried out at 700ºC for 10 hours and cooled in air with the rate < 10°C/hour. For preparing the $Na_{0.75}CoO_2$ (NCO075) film, the "sandwiched" $Co_3O_4$-(111) film was first muffled by NCO075 powder obtained by rapid heat-up sintering,[6] and then encapsulated in $Na_2CO_3$, as depicted schematically in Fig. 1 (specimen B). The pellet was then kept at 750ºC ~ 800ºC with oxygen flow for 5 hours. After the lateral diffusion of sodium, the $Co_3O_4$-(111) films evidently converted into NCO-(0001) films with the thickness expanding to ~ 250 nm.

Alternatively, Venimadhav et al.[7] has used $NaCOOCH_3$ powder to cover

Co$_3$O$_4$-(111) films in the annealing process to obtain NCO075 films. However, there were some unsettled issues concerning the uniformity and surface quality of the films obtained by that scheme. We also had tried other process schemes to obtain specimens with $x > 0.7$. For instance, the pellet of specimen A was annealed at 800ºC for few hours then quenched immediately in liquid nitrogen from 800ºC to keep as much sodium in the film as possible. However, the obtained films, though evidently displayed characteristics of NCO075, usually exhibited large residual resistivity indicating that there are more defects in the samples presumably resulted from quenching in liquid nitrogen. In contrast, with the specimen B scheme depicted in Fig. 1, we have been able to obtain NCO075 films reproducibly with very good uniformity and surface conditions. It is suggestive from the apparent concentration of the quenched samples that, even for the scheme of specimen A, the equilibrium sodium concentration at high temperature may have reached $x \geq 0.75$ with the encapsulation of Na$_2$CO$_3$ powders. The slow cooling practiced in scheme A, albeit being necessary for small residual resistivity, inevitably loses some sodium and finally arrives at the stable concentration of $x = 0.68$.

The crystal structure and sodium concentration of NCO thin films were carefully examined by x-ray diffraction (XRD) measurements. The results of $\theta$-$2\theta$ and $\Phi$ scans are shown in Fig. 2 and Fig. 3, respectively. The lattice mismatch between the NCO films ($a$ = 2.8407(2) Å for specimen A and $a$ = 2.843(1) Å for specimen B) and the sapphire substrate ($a$ = 4.760 Å) is reduced down to ~ 3% by a 30° in-plane rotation with respect to sapphire-($1\bar{1}00$), as being illustrated in the $\Phi$ scan results of Fig. 3. The results of Figs. 2 and 3 show that $Co_3O_4$-(111) films were grown epitaxially on sapphire-(0001) substrates, similar to the results obtained by Ohta *et al.*[4] and Venimadhav *et al.*[7]. The NCO films obtained by both lateral diffusion schemes also display highly epitaxial to the substrate, albeit becoming (0001)-oriented. Since in NCO cobaltates there exists an intimate correlation between the crystal structure parameters and Na content in the material, we inferred the Na content of the obtained NCO films by comparing the lattice constants with the structure phase diagram.[8] From the *c*-axis lattice constants of specimen A ($c$ = 10.9328(8) Å) and B ($c$ = 10.877(3) Å), the inferred sodium concentration for A and B are $x$ = 0.68 and 0.75, respectively. In addition, we noted that the NCO075 films instantly react with carbon dioxide and moisture in the ambient environment resulting in loss of sodium. This behavior is reflecting in the XRD results shown in Figs. 2 (c) and 2(d), in which there are still traces of $Co_3O_4$-(111) phase. As another indication of sodium loss, $\rho_{ab}(T)$ (Fig. 4(a)) where a bended $\rho$-$T$ curve characterizing NCO075 turned into a linear-like curve for NCO068 accompanied by a larger residual resistivity after a short period of exposure in air.

The $\rho_{ab}(T)$ measurements were carried out down to 0.4 K by the four-probes method. The $\rho_{ab}(T)$ curves shown in Fig. 4 display exactly identical temperature dependent behavior to that obtained from single crystals[2] with similar sodium contents of $x = 0.68$ and 0.75, respectively, indicating the consistency of inferring the film composition by the structural and transport property comparisons. Our films, notably, show an even smaller residual resistivity with the residual resistivity ratio $RRR \equiv \rho(300\text{ K})/\rho(0.4\text{ K}) = 37$ and 145, which are comparable or better than the $RRR$ of the single crystals in Ref. 2, for $x = 0.68$ and 0.75, respectively. In addition, the $\rho_{ab}$-$T$ of hydrolyzed NCO075 is illustrated in a gray dash line, which shows the characteristic of lower sodium concentration ($x < 0.75$) and larger residual resistivity. The in-plane Seebeck coefficient $S(T)$ represents yet another prominent property of this class of materials. We used a calibrated chip resistor glued to one end of the thin film as a pulsed-heat source. The temperature gradient was measured by using a differential thermocouple, and the corresponding Seebeck voltage was measured between the two signal leads connected by a pair of copper wires. The up-left inset of Fig. 4(a) shows that $S(300\text{ K})$ is nearly 80 μV/K and is strongly temperature dependent for the NCO068, consistent with the results of single crystals in the previous literature.[9] Unfortunately, due to the rapid deteriorating effects mentioned above, $S(T)$ for NCO075 thin films has not been available yet. In any case, the results presented above indicate that the preparation schemes in this letter can indeed produce films with very high quality.

Finally, we discuss briefly about the implications of low residual resistivity at low temperatures. For this purpose, infrared spectroscopy has been proved to be the effective tool for the analysis of the transport properties of a broad range of conducting systems.[10] Figure 4(b) shows the temperature dependent behavior of the far-infrared conductivity spectra for the NCO068 film. At 300 K, the low-frequency conductivity for $\omega < 100$ cm$^{-1}$ can be described by a Drude peak, which grows in intensity and sharpens as the temperature is lowered. By fitting the Drude conductivity and the measured infrared reflectance self-consistently, a nearly temperature-independent Drude plasma frequency $\omega_{pD} \sim 2400$ cm$^{-1}$ is obtained, whereas the scattering rate $1/\tau_D$ monotonically decreases with decreasing temperature (see the inset of Fig. 4(b)). Such behavior is typical of conventional metals.[11] Notably, the $1/\tau_D(20\ K)$ is about 6 cm$^{-1}$ compared with $1/\tau_D(300\ K) = 68$ cm$^{-1}$, consistent with the small residual resistivity (or large $RRR$). Since the residual resistivity and $1/\tau_D(20\ K)$ are proportional to the concentration of impurities or defects in the film, the transport and optical results indicate high quality of the NCO068 thin film.

In summary, NCO thin films with $x = 0.68$ and 0.75 were fabricated via R-SPE.[4] The thin films quality is, at least as decent as that of single crystals, judging from the results of XRD, $\rho_{ab}(T)$, and $S(T)$ curves. The environmental control of different equilibrium Na vapor pressures are achieved *reproducibly* by the present encapsulation schemes, which presumably have been very effective in providing sodium needed for forming NCO and at the same time preventing moisture from getting in to the films.

This work was supported by MOE ATU program and the National Science Council of Taiwan, under grants: NSC 95-2112-M-009-038-MY3, 95-2112-M-009-036-MY3, and 95-2112-M-009-035-MY3.

**Figure captions:**

**Figure 1:**

Schematics of the encapsulation schemes for preparing NCO thin films with $x = 0.68$ (specimen A) and 0.75 (specimen B). The diameter of pellets is 20 mm, and the size of substrates is about $5\times5$ mm$^2$.

**Figure 2:**

The XRD $\theta$-$2\theta$ scans of the as-grown samples of (a) Co$_3$O$_4$, (b) NCO068, and (c) NCO075 films. (d) was measured after exposing the NCO075 film in the ambient environment at $T = 25°C$ and humidity 42% for 1 hour. The symbols (+) and (*) label the sapphire substrate and sodium hydroxide hydrate (NaOH·$y$H$_2$O) peaks, respectively. The * labeled peaks of NaOH·$y$H$_2$O can be indexed from the data base, but $y$ is still undetermined.

**Figure 3:**

$\Phi$-scans of the $(1\bar{1}04)$ peak of the NCO films and the sapphire substrate indicate the relative rotation as well as the excellent epitaxial relations between films and substrate.

**Figure 4:**

(a) In-plane resistivity $\rho_{ab}$ versus $T$ curves (solid lines) of NCO thin films with $x = 068$ and 0.75. The gray dash line illustrates the $\rho_{ab}$-$T$ of hydrolyzed NCO075 in Fig. 2(d). Upper inset: $S(T)$ measurement for NCO068. Lower inset: the atomic force microscopic image (5×5 μm$^2$ and RMS = 1.67 nm) of NCO068 thin films was measured after thermal-diffusion process. (b) The temperature dependence of the far-infrared conductivity of the NCO068 thin film. The inset shows the temperature dependence of the Drude scattering rate $1/\tau_D$.

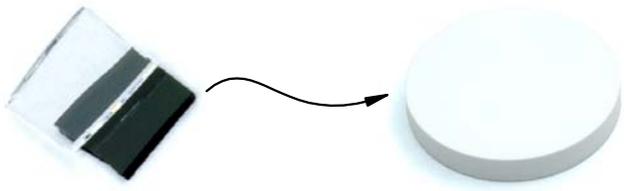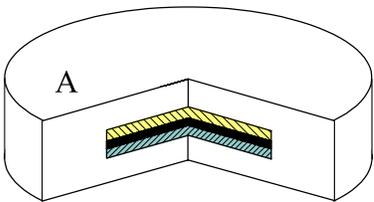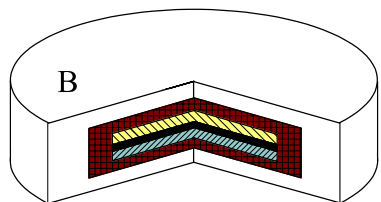

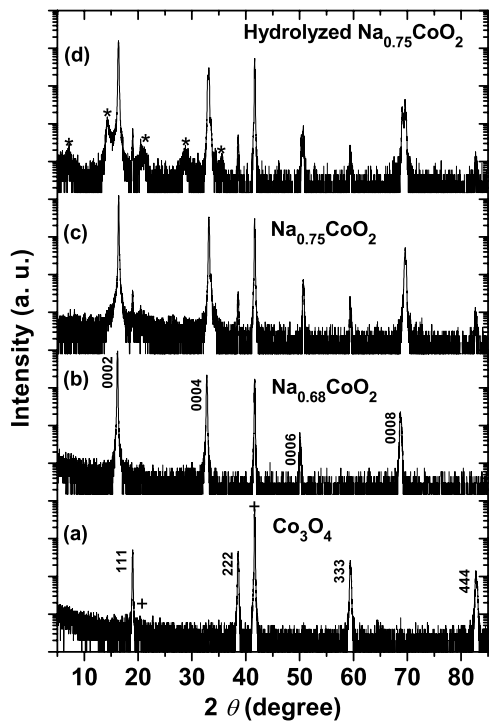

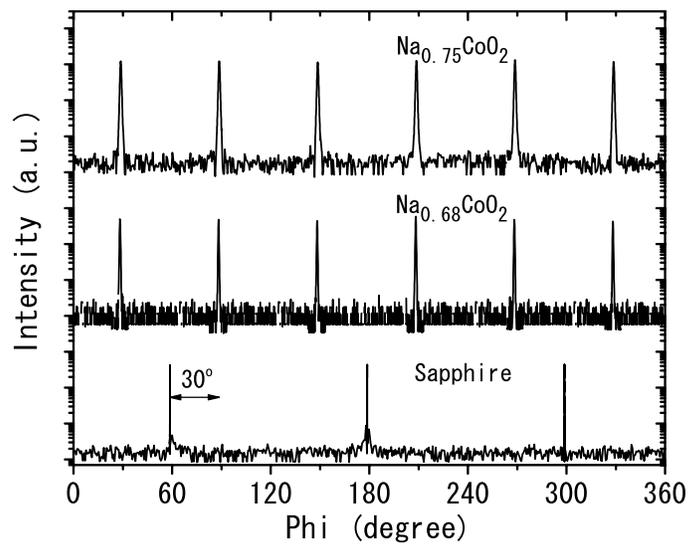

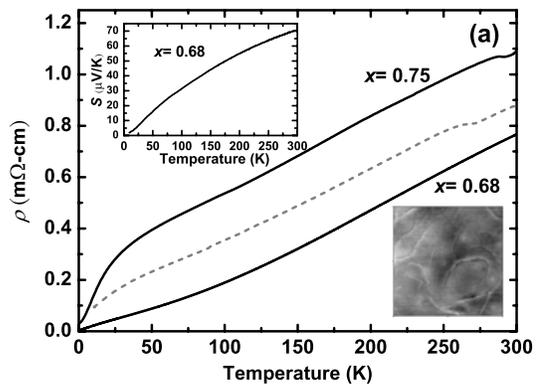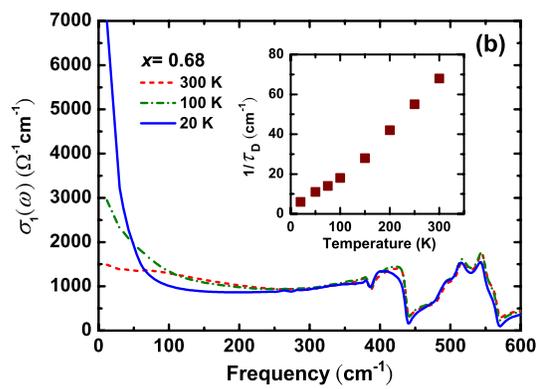